\documentclass[11pt]{article}
\usepackage[margin=1in]{geometry}
\usepackage{hyperref}
\usepackage{enumitem}
\usepackage{amsmath,amssymb,amsthm}
\usepackage{mathtools}
\usepackage{graphicx}
\usepackage{xcolor}
\usepackage{listings}
\usepackage{float}
\usepackage[T1]{fontenc}
\usepackage[utf8]{inputenc}
\usepackage{inconsolata}
\usepackage{upquote}
\usepackage{xpatch}
\usepackage{realboxes}
\usepackage{textcomp}
\usepackage[most]{tcolorbox}
\tcbuselibrary{breakable}
\usepackage{caption}
\usepackage{cleveref}

\usepackage{tikz}
\usetikzlibrary{arrows.meta,positioning,calc}

\definecolor{mycolor}{RGB}{0,88,204}
\definecolor{keywordcolor}{rgb}{0.7,0.1,0.1}
\definecolor{tacticcolor}{rgb}{0.0,0.1,0.6}
\definecolor{commentcolor}{rgb}{0.4,0.4,0.4}
\definecolor{symbolcolor}{rgb}{0.0,0.1,0.6}
\definecolor{sortcolor}{rgb}{0.1,0.5,0.1}
\definecolor{attributecolor}{rgb}{0.7,0.1,0.1}
\definecolor{mycolorSubtle}{RGB}{245,250,255}
\definecolor{darkgreen}{RGB}{0,100,0}
\definecolor{darkred}{RGB}{153,0,0}

\DeclareRobustCommand{\myinline}{\lstinline}
\makeatletter
\xpretocmd\myinline{\Colorbox{mycolorSubtle}\bgroup\appto\lst@DeInit{\egroup}}{}{}
\makeatother

\lstnewenvironment{codeLong}[1][]%
{
  \lstset{
    frame=none,
    language=lean,
    aboveskip=0mm,
    belowskip=2mm,
    xleftmargin=1mm,
    basicstyle=\ttfamily\scriptsize,
    columns=fixed,
    fontadjust=true,
    breaklines=true,
    escapeinside={(*}{*)},
    #1
  }
}
{}

\newcommand{\physLibDocsLink}[2][]{%
  \def\physLibTemp{#1}%
  \ifx\physLibTemp\empty
    \href{https://physlean.com/docs/find/?pattern=#2\#doc}{\texttt{#2}}%
  \else
    \href{https://physlean.com/docs/find/?pattern=#2\#doc}{\texttt{#1}}%
  \fi
}

\DeclareCaptionType{snippet}[Snippet][List of Snippets]
\crefformat{snippet}{\{\textcolor{darkgreen}{Input~#2#1#3}\}}
\Crefformat{snippet}{\textcolor{darkgreen}{\{#2#1#3\}}}

\newtcolorbox{startbox}{
  breakable,
  enhanced jigsaw,
  colback=gray!10,
  colframe=darkgreen,
  arc=2mm,
  boxrule=0.8pt,
  left=2mm, right=2mm, top=2mm, bottom=2mm,
  boxsep=0mm,
  fonttitle=\bfseries
}

\NewDocumentEnvironment{inputbox}{m}
{
  \begin{startbox}
  \captionof{snippet}{}
  \label{#1}
  \vspace{-0.25cm}
  {\color{darkgreen}\hrule} \vspace{2mm}
}
{
  \end{startbox}
}

\NewDocumentEnvironment{inputboxFoot}{}
{
  {\color{darkgreen}\hrule} \vspace{2mm}
}
{}

\newtcolorbox{derbox}{
  breakable,
  enhanced jigsaw,
  colback=gray!10,
  colframe=darkred,
  arc=2mm,
  boxrule=0.8pt,
  left=2mm, right=2mm, top=2mm, bottom=2mm,
  boxsep=0mm,
  fonttitle=\bfseries
}

\NewDocumentEnvironment{derboxFoot}{}
{
  {\color{darkred}\hrule} \vspace{2mm}
}
{}

\title{Physics as Code: From Scans to Theorems with ITP APIs in $SU(5)$ Model Building}
\author{
Sven Krippendorf$^{*}$ \quad Joseph Tooby-Smith$^{\dagger}$\\[0.75em]
\small $^{*}$ DAMTP and Cavendish Laboratory, University of Cambridge\\
\small Wilberforce Road, CB3 0WA, Cambridge\\
\small $^{\dagger}$ Department of Computer Science, University of Bath, Bath BA2 7AU, UK
}
\date{}

\newtheorem{theorem}{Theorem}

\begin{document}
\maketitle

\begin{abstract}
A recurring challenge in theoretical physics is to make reliable global statements about bounded but combinatorially large model spaces. Exhaustive scans quickly become opaque or impractical, while statistical exploration does not by itself provide theorem-backed guarantees. This motivates workflows in which the model-building problem itself is formalized inside an interactive theorem prover (ITP).

In this paper we develop an API-based methodology for formalizing such bounded model-building questions inside Lean, an interactive theorem prover. The central step is to represent the relevant charge spectra, predicates, and reduction moves as reusable ITP definitions, and then to derive the classification from proved reduction theorems rather than from an ad hoc scan. We demonstrate the strategy in a concrete \(SU(5)\) case study motivated by F-theory model building with additional Abelian symmetries.

At the charge-spectrum layer, we classify bounded spectra that admit a top-quark Yukawa coupling, avoid a selected set of dangerous operators, and satisfy a minimal charge-spectrum completeness condition. Our main result shows that every such spectrum in the bounded search space arises from finitely many minimal top-Yukawa witnesses together with controlled completions and certified closure steps. This classification represents a formally verified description of the full viable class in the charge-spectrum setting studied here.

The development is implemented inside PhysLib as reusable infrastructure rather than as a one-off verification script. It provides a proof of principle for how interactive theorem provers can turn combinatorially difficult model-building problems into correctness-first, reusable workflows, and we discuss how the resulting certified classification can serve as reliable input for downstream analyses.
\end{abstract}

\newpage
\tableofcontents
\newpage

\section{Introduction}

\paragraph{The larger scientific goal.}
A central long-term goal of string phenomenology is to understand whether string theory can provide a viable unified description of gravity and the other fundamental forces that is consistent with experimental constraints. Addressing that question requires statements about whole \emph{classes} of low-energy effective field theories rather than isolated examples. One would like to know whether a given construction admits phenomenologically viable vacua at all, whether certain classes are excluded, how viability depends on assumptions, and how surviving models are distributed across a bounded region of model space. This viewpoint lies naturally between the \emph{string landscape}, which aims to describe the space of consistent vacua \cite{Susskind:2003kw,Douglas:2003um,Acharya:2006zw}, and the \emph{swampland programme}, which seeks criteria separating effective theories that can arise from quantum gravity from those that cannot \cite{Ooguri:2006in,Palti:2019pca}.

In practice, however, progress is often driven by explicit constructions, bounded scans, and statistical or sampling-based models. These tools are valuable, but their conclusions inherit the limits of the assumptions and guarantees that accompany them. In some settings rigorous global statements are possible --- for example, under suitable assumptions one can prove finiteness statements for parts of the flux landscape \cite{Grimm:2023lrf}. More broadly, one is still faced with large combinatorial model spaces for which case-by-case analysis, sampling, or statistical surrogates do not by themselves provide the desired level of control. This has motivated substantial work based on counting, bounded enumeration, and geometric bounds \cite{Loges:2022mao,Gendler:2023ujl,Chandra:2023afu}. Our aim is different: not merely to certify a final list of outputs, but to formalize and certify the reduction mechanism that produces that list inside an interactive theorem prover. The same tension appears well beyond string phenomenology, across beyond-the-Standard-Model settings where gauge sectors, charge assignments, couplings, and consistency conditions proliferate combinatorially.

\paragraph{A representative example: GUT model search with additional Abelian symmetries.}
A representative example, and the one used throughout this paper, is \(SU(5)\) model building with additional Abelian symmetries, following the notation and phenomenological setup of \cite{Krippendorf:2015kta} and related F-theory constructions such as \cite{Dudas:2009hu,Krippendorf:2014xba}. Even in this comparatively concrete setting, one must choose charges, determine which operators are allowed or forbidden, and impose a minimal charge-spectrum completeness condition requiring both Higgs sectors and both matter sectors to be present. Historically, single-\(U(1)\) models and a few classes with two additional \(U(1)\) symmetries could still be analysed systematically, but beyond that the combinatorial growth rapidly becomes prohibitive. We use this setting as a proxy for a broader issue: how to turn a bounded search over many candidate models into a structural classification with explicit guarantees, rather than an implementation-dependent scan of isolated examples.

\paragraph{Factoring out the combinatorics.}
The key idea is to prove statements that shrink the relevant search problem before any explicit enumeration takes place. Here ``combinatorics'' refers to the rapid growth of candidate charge spectra together with the operators, exclusions, and completion moves that must be checked. The point is not merely to scan faster, but to identify reusable structure that removes large parts of the search space \emph{a priori}. In the present context, \emph{monotonicity} means that certain predicates behave predictably under enlargement of a charge spectrum; for example, once a coupling is present, it remains present in any larger spectrum containing the relevant charges. \emph{Closure} means that after one has found a certified seed or certified completion, the admissible enlargement moves used in the argument remain inside a controlled candidate class. \emph{Scaling} therefore no longer means spending more compute on a larger scan. It means proving the reduction once and reusing it as the bounded class is enlarged. Our suggestion is to build this logic directly into the model-building workflow in a reusable formal form.

\paragraph{The proposed mechanism: theorem-backed APIs in an interactive theorem prover.}
We implement this viewpoint by treating the physics subproblem as an \emph{application programming interface} (API) design problem inside Lean, an interactive theorem prover \cite{de2015lean}. The basic object of the API makes the physical data explicit, and the interface definitions encode the questions physicists actually ask --- whether a term is allowed, whether a spectrum is complete, whether dangerous operators appear, and so on. On top of this interface one proves reusable lemmas and \emph{reduction theorems}, by which we mean theorems showing that every object satisfying the target predicates can be generated from smaller certified witnesses by controlled completion and enlargement steps. The point is not to replace the usual physics language, but to recast it in a form precise enough to support proof reuse.

This perspective builds on the emerging use of Lean and library-based formalization in high-energy physics and related areas \cite{Tooby-Smith:2024vqu,PhysLibRepo,LeanQuantumInfoRepo, JTSPerspective}. The deliverable is therefore not only a proof-checked argument but a reusable semantic layer that later formalizations can extend without rebuilding the whole reasoning stack. That is the route to scalability we advocate: invest once in stable definitions and reductions, then reuse them across larger bounded classes and related model-building problems.
We also note recent formalization efforts in physics and neighbouring domains, including quantum information~\cite{Meiburg:2025mwn}, formal QFT~\cite{Douglas:2026hyk}, agentic autoformalization for quantum computation~\cite{Ren:2026tdj}, and in physical chemistry~\cite{Josephson:2025chemrxiv}.

\paragraph{What we demonstrate in this paper.}
We instantiate this programme in a concrete string-inspired \(SU(5)\) setting with additional Abelian symmetries, mentioned above, closely following the notation and operator language of \cite{Krippendorf:2015kta} within the broader F-theory GUT context of \cite{Dudas:2009hu,Krippendorf:2014xba}. At the charge-spectrum layer, the bounded problem is to classify those spectra that admit a top Yukawa coupling, forbid a selected set of dangerous superpotential and K\"ahler potential operators, and satisfy a minimal completeness condition requiring both Higgs sectors and both matter sectors to be present. Separately, we track one-step Yukawa-induced regeneration only in the superpotential sector. The example is intended as a proof of principle for the formal workflow rather than as a complete phenomenological analysis, but it is already large enough to show how the reduction changes the problem from an opaque search over raw candidates to a theorem-backed classification of the viable charge spectra.

\begin{center}
\fbox{%
\begin{minipage}{0.94\linewidth}
\textbf{Main result.}
Within the bounded class studied here, every viable complete model arises from finitely many minimal top-Yukawa witnesses together with controlled completions. In this sense, the theorem replaces exhaustive generation over a combinatorially exploding ambient class by a certified construction of exactly the viable complete models.
\end{minipage}}
\end{center}

This has an immediate practical consequence. Later phenomenological refinements can be applied to a theorem-backed reduced class rather than to an intractably large raw search space. The development is implemented as a reusable Lean component inside PhysLib, so the main point is not only one classification theorem, but a reusable formal interface that can guide future formalizations in larger charge spaces and related beyond-the-Standard-Model settings.

The rest of the paper is organized as follows. Section~\ref{sec:physics-to-formal} stays at the physics level: it fixes the running \(SU(5)\) example, formulates the bounded classification problem, and motivates the reduction strategy without yet introducing Lean definitions. Section~\ref{sec:formal-vocabulary} then translates those ingredients into a reusable PhysLib API. Section~\ref{sec:certified-reduction} states the certified reduction theorem and explains how the proof is assembled from witness, completion, and closure lemmas. Section~\ref{sec:executable-classification} turns that certified reduction into a concrete computation and clarifies what is proved versus what is merely evaluated. We then summarize lessons learned and outline possible extensions.

\begin{figure}[t!]
\centering
\begin{tikzpicture}[
  x=1cm,y=1cm,
  >=Latex,
  font=\small,
  line/.style={-Latex, thick, draw=black!75},
  physics/.style={
    draw=black!65,
    rounded corners=2mm,
    thick,
    fill=blue!6,
    align=center,
    text width=4.20cm,
    minimum height=1.55cm,
    inner sep=5pt
  },
  formal/.style={
    draw=black!65,
    rounded corners=2mm,
    thick,
    fill=orange!8,
    align=center,
    text width=3.55cm,
    minimum height=1.20cm,
    inner sep=5pt
  },
  process/.style={
    draw=black!65,
    rounded corners=2mm,
    thick,
    fill=gray!10,
    align=center,
    text width=6.45cm,
    minimum height=1.25cm,
    inner sep=6pt
  },
  theorem/.style={
    draw=black!65,
    rounded corners=2mm,
    thick,
    fill=green!8,
    align=left,
    text width=3.8cm,
    inner sep=6pt
  },
  outputbox/.style={
    draw=black!65,
    rounded corners=2mm,
    thick,
    fill=green!5,
    minimum width=11.25cm,
    minimum height=5.0cm,
    inner sep=8pt
  }
]

\node[process] (question) at (0,0) {%
  \textbf{Physical model-building question}\\[0.8mm]
  Which bounded charge spectra are phenomenologically viable?
};

\node[physics, anchor=north] (class) at (-2.70,-1.70) {%
  \textbf{Model class \(\mathcal C\)}\\[0.8mm]
  \(SU(5)\) charge spectra with additional \(U(1)\) data
};

\node[physics, anchor=north] (props) at (2.70,-1.70) {%
  \textbf{Physics requirements \(\mathcal P\)}\\[0.8mm]
  top Yukawa present, dangerous operators absent
};

\node[formal, anchor=north] (core) at (-2.70,-4.25) {%
  \textbf{Lean core object}\\[0.8mm]
  \texttt{ChargeSpectrum}
};

\node[formal, anchor=north] (preds) at (2.70,-4.25) {%
  \textbf{Lean predicates}\\[0.8mm]
  {\scriptsize\texttt{AllowsTerm}, \texttt{IsComplete},}\\
  {\scriptsize Yukawa-regeneration predicates}
};

\node[process] (problem) at (0,-7.55) {%
  \textbf{Formalized classification problem in Lean}\\[0.8mm]
  classify those \(x \in \mathcal U(I)\) that satisfy the target predicates
};

\node[outputbox] (output) at (0,-12.15) {};
\node[font=\bfseries] at ($(output.north)+(0,-0.34)$) {Certified classification output};

\coordinate (setcenter) at ($(output.center)+(-2.45,-0.10)$);

\draw[thick] (setcenter) ellipse [x radius=2.95cm, y radius=1.85cm];
\draw[thick, fill=green!12] (setcenter) ellipse [x radius=2.20cm, y radius=0.82cm];

\node[anchor=west, font=\footnotesize] at ($(setcenter)+(-1.85,1.15)$) {%
  \(\mathcal U(I)\): bounded model class
};

\node[anchor=center, font=\scriptsize\bfseries] at ($(setcenter)+(0.18,0.00)$) {%
  \(\mathcal V(I)\): certified viable class
};

\node[theorem] (thm) at ($(output.center)+(2.95,0.02)$) {%
  \textbf{Provable endpoint}\\[0.8mm]
  \emph{Soundness:} if \(x \in \mathcal V(I)\), then \(x\) is viable and complete.\\[1.0mm]
  \emph{Completeness:} every viable complete \(x \in \mathcal U(I)\) lies in \(\mathcal V(I)\).
};

\draw[line] (question) -- (class.north);
\draw[line] (question) -- (props.north);
\draw[line] (class.south) -- (core.north);
\draw[line] (props.south) -- (preds.north);
\draw[line] (core.south) -- (problem);
\draw[line] (preds.south) -- (problem);
\draw[line] (problem) -- (output);

\end{tikzpicture}
\caption{
Workflow of the certified classification. A physical model-building question is decomposed into a model class \(\mathcal C\) and a set of physics requirements \(\mathcal P\). Both are formalized in Lean: the model class through the core object \myinline{ChargeSpectrum}, and the target requirements through Lean predicates. This yields a formal classification problem on the bounded class \(\mathcal U(I)\), whose output is the certified viable class \(\mathcal V(I)\), together with soundness and completeness guarantees.
}
\label{fig:workflow-certified-classification}
\end{figure}
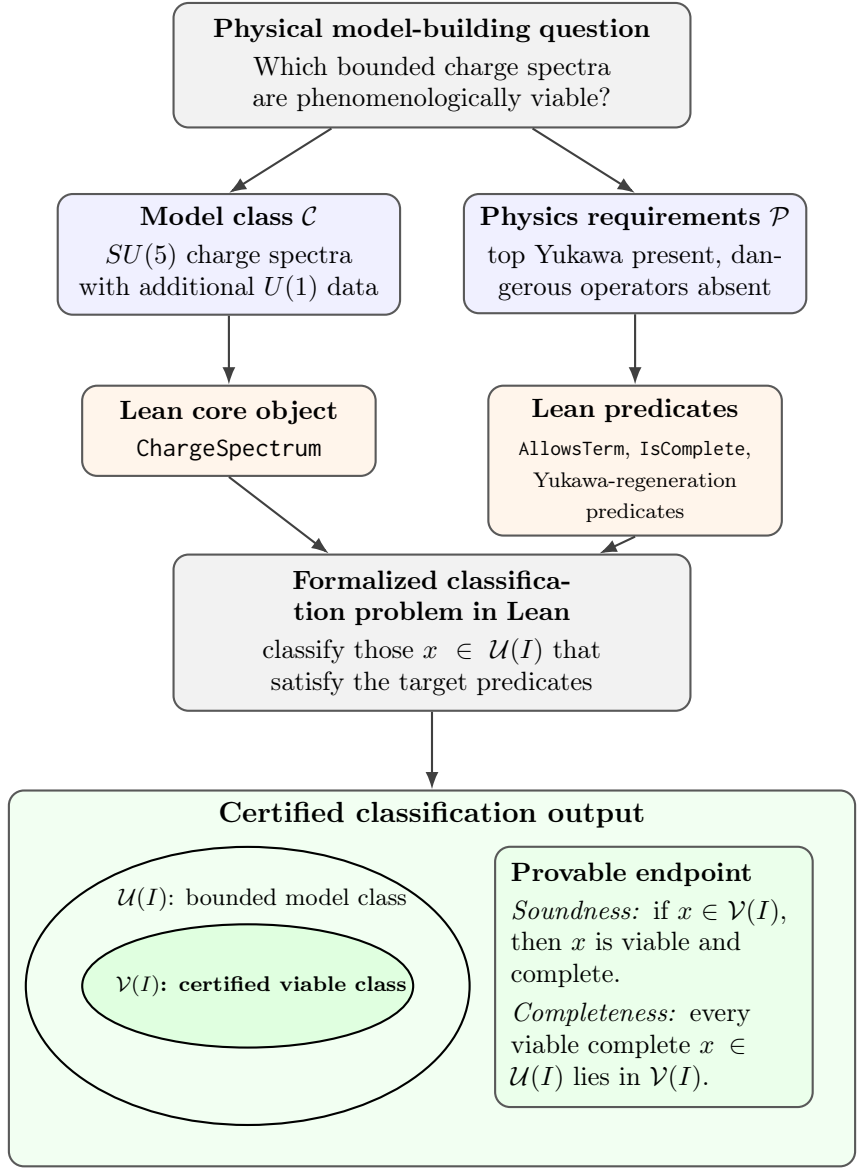

Figure~\ref{fig:workflow-certified-classification} gives a roadmap for the certified workflow developed in the remainder of the paper. The code itself can be found at 
\begin{center}
\url{https://github.com/leanprover-community/physlib}.
\end{center}

\section{From the Physics Question to a Formal Classification Problem}
\label{sec:physics-to-formal}

The purpose of this section is to bridge from the informal physics question to the formal classification problem later implemented in PhysLib. Although the motivating example comes from supersymmetric \(SU(5)\) model building with additional Abelian symmetries, the reduction pattern itself is more general and not tied only to string-specific input.

The target classification problem is to characterize, within a bounded search space of charge spectra, exactly those spectra that admit the top Yukawa coupling, avoid the selected dangerous operators, avoid one-step Yukawa-induced regeneration in the superpotential sector, and satisfy the minimal charge-spectrum completeness condition.

\subsection{The running example and its physics origin}

In this subsection we follow the notation and phenomenological setup of \cite{Krippendorf:2015kta} within the broader F-theory context of \cite{Dudas:2009hu,Krippendorf:2014xba}. In that literature, one considers charges for matter multiplets in the \(\mathbf{10}\) and \(\mathbf{\bar 5}\) representations together with charges for the Higgs multiplets, and asks which charge assignments are compatible with a chosen set of phenomenological requirements. In the present paper the broad question is:
\begin{quote}
Given a bounded set of allowed charges, which charge spectra can give rise to phenomenologically viable models?
\end{quote}
This style of question is familiar from earlier model-building work: one imposes selection-rule constraints, requires a top Yukawa sector, forbids a chosen list of dangerous operators, and studies the surviving charge assignments.

At this stage it is useful to say informally what we mean by a \emph{charge spectrum}. For us, a charge spectrum records which distinct Higgs and matter charges occur in the relevant sectors. At this level it does not record multiplicities of curves carrying the same charge.
A convenient toy spectrum at the level used in this paper is simply the record that the charge \(-1\) occurs in the \(\mathbf{10}\)-sector, the charge \(0\) occurs in the \(\overline{\mathbf 5}\)-sector, and the Higgs charges are \(q(H_u)=-2\) and \(q(H_d)=-3\). It should be read as a record of distinct charge values present in each sector, not as a full multiplicity assignment. This toy example is deliberately too small to be realistic, but it already illustrates the finite charged data that later becomes the formal object.

The present paper focuses on the \emph{charge-spectrum layer} of the problem. A full phenomenological pipeline must later incorporate further ingredients such as flux data, chirality assignments, anomaly cancellation, and conditions such as the absence of exotics. The point here is not to address those later stages, but to isolate one layer on which a mathematically controlled theorem can already be proved. This is enough to demonstrate the workflow while keeping the first object conceptually clean.

For the notation and operator language used below, we stay close to \cite{Krippendorf:2015kta} and the related F-theory discussion in \cite{Dudas:2009hu}. At the level treated in this paper, the formal object records only the spectrum of distinct charges, not multiplicities. The phenomenological exclusion imposed at the charge-spectrum level is chosen to match the Lean predicate \myinline{IsPhenoConstrained}. Concretely, we forbid the dangerous superpotential operators
\begin{align}
  & \mu\, 5_{H_u}\,\bar 5_{H_d}, \\
  & \beta_i\, \bar 5_M^i\,5_{H_u}, \\
  & \Lambda_{ijk}\,\bar 5_M^i\,\bar 5_M^j\,10^k, \\
  & W^1_{ijkl}\,10^i\,10^j\,10^k\,\bar 5_M^l, \\
  & W^2_{ijk}\,10^i\,10^j\,10^k\,\bar 5_{H_d}, \\
  & W^4_i\,\bar 5_M^i\,\bar 5_{H_d}\,5_{H_u}\,5_{H_u},
\end{align}
and, following Eq.~(3.12) of \cite{Dudas:2009hu}, also the dangerous K\"ahler-potential operators
\begin{align}
  & K^1_{ijk}\,10_M^i\,10_M^j\,5_M^k, \\
  & K^2_i\,\bar 5_{H_u}\,\bar 5_{H_d}\,10_M^i .
\end{align}
These eight operator labels are exactly the ones later bundled into \myinline{IsPhenoConstrained}. The regeneration analysis is a separate ingredient: here we only test whether dangerous \emph{superpotential} operators can be re-generated, up to the chosen level, by Yukawa-induced singlet insertions. Later, in the formal API, all operator labels --- including \myinline{W3}, \myinline{topYukawa}, and \myinline{bottomYukawa} --- are packaged into a finite datatype of potential-term labels.

A simple bounded example to keep in mind is the toy charge menu above: even there one already has to choose optional Higgs entries and subsets of allowed matter charges, and the combinatorics become much worse once the charge menus grow.

\subsection{Why brute-force scans are not enough for a certified classification}

A conceptually straightforward baseline is to enumerate all charge assignments in a bounded model class and then filter them by the desired constraints. This quickly becomes unsatisfactory for two reasons.
\begin{itemize}
  \item \textbf{Conceptual opacity and implementation dependence.} Even in a finite search space, the correctness of a brute-force algorithm can depend on implementation details: how spectra are generated, which branches are pruned, whether symmetries are quotiented correctly, and whether all relevant cases are actually reached. A raw scan does not by itself make these points transparent.
  \item \textbf{Poor scaling of the search space.} As the allowed charge menu grows, the number of candidate spectra and associated checks grows rapidly. What is manageable in a tiny example becomes opaque or infeasible in a broader bounded class.
\end{itemize}
These combinatorial effects become even more severe once the goal is not just to search, but to prove an exhaustive classification.
Historically, one could still make fairly explicit global statements for very small classes, in particular for single-\(U(1)\) models, and with substantial effort some classes with two additional \(U(1)\) symmetries could also be treated. Beyond that, however, the combinatorics become too severe for direct scanning to remain a satisfactory route to global statements.

A theorem-backed classification has a different scientific status from an opaque scan. Exclusion from the final class is then not merely the outcome of a script, but the consequence of a proved statement that every spectrum satisfying the target conjunction arises from certified witnesses and completions inside the bounded search space. This is also where scaling changes character: one scales by reusing the proved reduction, not by rerunning a larger opaque enumeration.

\subsection{Minimal witnesses and controlled completions}

The main reduction strategy follows a very physical way of thinking about model building. In practice, one rarely starts from the full ambient search space of all possible models. Instead, one first identifies the minimal ingredients that realize the local structure one cares about, and then asks how these ingredients can be completed into full models without spoiling the desired properties. Our proof strategy mirrors exactly this intuition.

First, we isolate \emph{minimal witnesses}: small charge spectra that already realize the key local structure of interest. In the present paper this means minimal spectra that allow the top Yukawa term. Here ``minimal'' should be understood literally: a witness is minimal if it allows the top Yukawa coupling, but no proper sub-spectrum does. Equivalently, one cannot remove any charged ingredient from the witness and still retain the desired top-Yukawa structure.

In the toy example above, suppose the top Yukawa is realized by a neutral coupling of the form
\[
  10_{-1}\,10_{-1}\,H_u .
\]
Here \(10_q\) denotes a \(\mathbf{10}\)-curve of charge \(q\), so in particular \(10_{-1}\) means a \(\mathbf{10}\)-curve of charge \(-1\). Then the spectrum consisting only of the charge \(-1\) for the relevant \(10\)-curve together with the Higgs-up charge \(-2\) is a minimal witness for the top Yukawa: if one removes either constituent, the coupling is no longer available.

Second, we study \emph{controlled completions}. Once such a minimal witness is fixed, one asks which additional Higgs and matter curves can be added so as to produce a complete model while preserving the relevant viability predicates. In the toy example, this means adding the further matter and Higgs sectors required for completeness --- for instance, adding the missing \(\bar 5\) matter sector and the down-type Higgs sector --- without introducing forbidden operators. In the actual theorem this completion step is not arbitrary: it is constrained by the same physics logic that guides ordinary model building.

This is an important point conceptually. The proof does not replace a physicist's way of thinking by something alien. Rather, it formalizes a familiar model-building strategy: identify the smallest local seed with the desired coupling structure, then enlarge it in controlled ways until a complete model is obtained. In this sense, the proof remains guided by physics intuition. This already previews the proof structure of the later proof: first isolate minimal witnesses, then characterize admissible completions, and finally prove closure and exhaustiveness in the bounded search space. For instance, one may think of the bounded one-\(U(1)\) and two-\(U(1)\) cases as concrete instances of the general witness-and-completion pattern.

There is also a suggestive analogy with generative models. A learned generative model starts from a restricted seed distribution and produces candidate outputs by a controlled generation mechanism. Here too the theorem identifies a finite family of admissible seeds --- the minimal witnesses --- and a controlled completion procedure that generates the candidate class. The crucial difference is that in the present setting the generation mechanism is theorem-backed and exhaustive within the bounded search space rather than merely statistical.

\subsection{A mathematical summary in PhysLib-style notation}

We now rewrite the discussion in a notation that is already close to the eventual formalization in PhysLib.

Let \(\mathcal Z\) denote the ambient charge type, and let
\[
  S_{\bar 5}, S_{10} \subset \mathcal Z
\]
be finite sets of allowed charges. These determine a bounded model class of charge spectra,
\[
  \mathcal U(S_{\bar 5},S_{10}) := \mathrm{ofFinset}\; S_{\bar 5}\; S_{10}.
\]
Concretely, this means that every Higgs and matter charge appearing in the spectrum must be drawn from the chosen bounded menus for the appropriate representation. Inside this bounded model class we want to characterize those spectra \(x\) satisfying the following target conjunction:
\begin{equation}
\label{eq:target-condition-summary}
\begin{aligned}
  &\mathrm{AllowsTerm}\;x \;\mathrm{topYukawa}\;\wedge \\
  &\neg\,\mathrm{IsPhenoConstrained}\;x \;\wedge \\
  &\neg\,\mathrm{YukawaGeneratesDangerousAtLevel}\;x\;1 \;\wedge \\
  &\mathrm{IsComplete}\;x.
\end{aligned}
\end{equation}
This conjunction defines the charge spectra that we ultimately want to classify.

The individual predicates have direct physical meanings:
\begin{itemize}
  \item \(\mathrm{AllowsTerm}\;x\;\mathrm{topYukawa}\) says that the charge spectrum \(x\) admits the top Yukawa coupling.
  \item \(\mathrm{IsPhenoConstrained}\;x\) bundles the selected dangerous operators introduced earlier and forbids them already at the charge-spectrum level.
  \item \(\mathrm{YukawaGeneratesDangerousAtLevel}\;x\;1\) says that a dangerous operator is re-generated after one Yukawa-induced singlet insertion.
  \item \(\mathrm{IsComplete}\;x\) encodes the minimal requirement that both Higgs sectors and both matter sectors are present.
\end{itemize}

It is also useful to make the operator regeneration step explicit. Suppose an operator \(\mathcal O\) carries total charge \(q(\mathcal O)\), and singlets \(S_k\) with charges \(q(S_k)\) acquire vacuum expectation values. If integers \(n_k \ge 0\) exist such that
\begin{equation}
\label{eq:regeneration-condition}
  q(\mathcal O) + \sum_k n_k\, q(S_k) = 0 ,
\end{equation}
then \(\mathcal O\) can be re-generated with suppression
\begin{equation}
\label{eq:regeneration-suppression}
  \prod_k \left(\frac{\langle S_k\rangle}{M_{\mathrm{GUT}}}\right)^{n_k}.
\end{equation}
For a single singlet this reduces to the standard Froggatt--Nielsen expression
\[
  \left(\frac{\langle S\rangle}{M_{\mathrm{GUT}}}\right)^n .
\]
In particular, ``level \(1\)'' means that one Yukawa-generated singlet insertion already suffices to regenerate a dangerous operator.

We have already referred informally to a reduction theorem; here is the precise meaning in the present setting. Rather than classifying all
\[
  x \in \mathcal U(S_{\bar 5},S_{10})
\]
satisfying the target conjunction in Eq.~\eqref{eq:target-condition-summary} by brute force, one proves that every such \(x\) arises from finitely many minimal witnesses together with controlled completions. In other words, the full viable class at this layer is generated from a finite seed class in a mathematically controlled and exhaustive manner.

The formal task is therefore to define these predicates precisely, prove the reduction statement, and then execute the resulting finite classification. The next sections explain how these definitions are encoded in PhysLib and how the proof is organized.

\section{Formal Vocabulary: a PhysLib API for Charge Spectra}
\label{sec:formal-vocabulary}

Section~\ref{sec:physics-to-formal} reformulated the physics question as a classification problem for the viability-and-completeness condition summarized in Eq.~\eqref{eq:target-condition-summary}. We now introduce the formal vocabulary in which that condition is stated inside PhysLib. This section is deliberately architectural: the goal is to build a reusable API, not yet to jump straight to the main theorem. The immediate case study is the \(SU(5)\) bounded charge-spectrum problem, but the interface is intentionally more general and can serve as a guide for future formalizations of other bounded model-building questions.

From the physics point of view, the goal is simple: we want a formal language that talks about exactly the same objects and questions that appear in ordinary model building. A charge spectrum should be an explicit object. Questions such as whether a term is allowed, whether a spectrum is complete, or whether it belongs to a bounded class should become reusable definitions. In that sense, the formalization does not replace the way physicists think; it rewrites that familiar reasoning in a form precise enough to support proofs.

Throughout this section it is helpful to keep in mind one tiny toy spectrum, as above. Let the charge type be \(\mathcal{Z}=\mathbb Z\), and consider a spectrum with the following data:
\[
q_{H_d}=-3,\qquad q_{H_u}=-2,\qquad Q_{\bar 5}=\{0\},\qquad Q_{10}=\{-1\}.
\]
In words, this toy spectrum contains a down-type Higgs sector of charge \(-3\), an up-type Higgs sector of charge \(-2\), one \(\overline{\mathbf 5}\)-matter charge \(0\), and one \(\mathbf{10}\)-matter charge \(-1\). We denote this spectrum by \(x_{\mathrm{toy}}\).

In the Lean-oriented notation used below, the same object is written as
\[
  x_{\mathrm{toy}} :=
  \langle \mathrm{some}\,(-3),\ \mathrm{some}\,(-2),\ \{0\},\ \{-1\}\rangle .
\]
Here \(\mathrm{some}\,q\) means that the corresponding Higgs charge is present and equal to \(q\), while \(\mathrm{none}\) would mean that the corresponding Higgs sector is absent. The sets \(\{0\}\) and \(\{-1\}\) are singleton finite sets containing exactly the indicated charges. We will occasionally use this notation because it mirrors the code directly.

\subsection{Charge spectra as the basic object}

The basic physical object in our story is a finite package of Higgs and matter charges. This is exactly what a physicist writes down when specifying a candidate charge assignment. The API is therefore centered on a single basic object, the charge spectrum. In the present case it records the optional Higgs charges and the finite sets of distinct matter charges in the \(\overline{\mathbf 5}\) and \(\mathbf{10}\) sectors. In the implementation used for this paper, this definition lives centrally in the charge-spectrum part of PhysLib, so that later formalizations can import the same conventions rather than restating them locally.

Two modelling choices are important.
\begin{itemize}
  \item The Higgs charges are stored as optional values, because a sector may be absent.
  \item The matter charges are stored as \texttt{Finset}s. A \myinline|Finset| is a finite set in Lean: it records which distinct charges occur, but not multiplicities. This matches the level of structure needed for the selection-rule questions treated here.
\end{itemize}
We also parameterize the object by a general charge type \(\mathcal{Z}\). This is important conceptually: the same formal interface can be used for one \(U(1)\), several \(U(1)\)s, or suitable discrete symmetry data, provided the required algebraic operations are available.

We now display the core Lean structure. Its type is \myinline|ChargeSpectrum 𝓩|, and its fields encode exactly the data just described in physics notation: optional Higgs charges and finite sets of distinct matter charges. We display the definition explicitly because later predicates and theorems refer back to these fields, and in Lean one can always chase such definitions back from a later goal to the underlying structure.

\begin{inputbox}{code:ChargeSpectrum}
\begin{codeLong}
structure (*\physLibDocsLink[ChargeSpectrum]{SuperSymmetry.SU5.ChargeSpectrum}*) (𝓩 : Type := ℤ) where
  /-- The charge of the `Hd` particle. -/
  qHd : Option 𝓩
  /-- The negative of the charge of the `Hu` particle. That is to say,
    the charge of the `Hu` when considered in the 5-bar representation. -/
  qHu : Option 𝓩
  /-- The finite set of charges of the matter fields in the `Q5` representation. -/
  Q5 : Finset 𝓩
  /-- The finite set of charges of the matter fields in the `Q10` representation. -/
  Q10 : Finset 𝓩
\end{codeLong}
\begin{inputboxFoot}
\footnotesize Structure of the basic PhysLib charge-spectrum object.
\end{inputboxFoot}
\end{inputbox}

This is the point where the physics conventions are made explicit. In particular, the implementation stores the up-type Higgs charge in the \(\overline{\mathbf 5}\) convention, so helper functions can recover the corresponding \(\mathbf 5\) charge by negation when required. That kind of small interface decision matters: once the convention is fixed centrally, the later definitions and proofs do not have to re-encode it.

The toy spectrum \(x_{\mathrm{toy}}\) above is an example of exactly this object type. It is nothing more mysterious than a finite charge assignment written as structured data. This is a good example of a general theme in the paper: the code-level object is simply the familiar physics object with its conventions made explicit.

\subsection{Structural relations and instances}

After fixing the basic object, the next step is to endow it with the elementary relations physicists already use informally. In model building, we often say that one spectrum is obtained from another by removing some sectors. The formalization should support exactly this language.

Two elementary structural notions are needed immediately. First, there is an empty charge spectrum, in which both Higgs sectors are absent and both matter sectors are empty. Second, there is a subset relation, expressing that one spectrum is obtained from another by deleting Higgs sectors or matter charges. These two notions provide the formal language for later statements about minimality and completion.

\begin{inputbox}{code:structuralInstances}
\begin{codeLong}
instance (*\physLibDocsLink[emptyInst]{SuperSymmetry.SU5.ChargeSpectrum.emptyInst}*) : EmptyCollection (ChargeSpectrum 𝓩) where
  emptyCollection := ⟨none, none, {}, {}⟩

instance (*\physLibDocsLink[hasSubset]{SuperSymmetry.SU5.ChargeSpectrum.hasSubset}*) HasSubset (ChargeSpectrum 𝓩) where
  Subset x y :=
    x.qHd.toFinset ⊆ y.qHd.toFinset ∧
    x.qHu.toFinset ⊆ y.qHu.toFinset ∧
    x.Q5 ⊆ y.Q5 ∧
    x.Q10 ⊆ y.Q10
\end{codeLong}
\begin{inputboxFoot}
\footnotesize Empty spectrum and subset relation for charge spectra.
\end{inputboxFoot}
\end{inputbox}

For example, let \(y_{\mathrm{toy}}\) be the spectrum obtained from \(x_{\mathrm{toy}}\) by removing the down-type Higgs sector while leaving the other data unchanged. In ordinary mathematical notation, this means
\[
q_{H_d}\ \text{absent},\qquad q_{H_u}=-2,\qquad Q_{\bar 5}=\{0\},\qquad Q_{10}=\{-1\}.
\]
In the Lean-oriented notation this is written as
\[
  y_{\mathrm{toy}} := \langle \mathrm{none},\ \mathrm{some}\,(-2),\ \{0\},\ \{-1\}\rangle .
\]
Thus \(y_{\mathrm{toy}} \subseteq x_{\mathrm{toy}}\): it is obtained from \(x_{\mathrm{toy}}\) by deleting one sector.

These instances do two things at once. They give the object the notation one expects mathematically --- in particular an empty spectrum and a subset relation --- and they connect the development to standard Lean and Mathlib proof patterns for set-like objects. This is already one place where the wider library starts to matter: once these structural instances are declared, later arguments can reuse generic notation, rewriting lemmas, and monotonicity patterns instead of reproving elementary facts for this one case. Much of the leverage of an API-based formalization begins precisely here.

\subsection{Interface definitions: the verbs of the API}

Once the basic object has been fixed, the next step is to formalize the questions physicists actually ask about it. This is really the natural way we think in model building: does a given charge assignment allow a desired coupling, does it forbid dangerous operators, is it complete, and can it be extended? The only difference is that these questions are now written as reusable definitions in code.

We will repeatedly use two kinds of interface definitions:
\begin{itemize}
  \item \textbf{data-valued definitions}, which return new objects such as bounded classes of charge spectra, powersets, multisets, or completion candidates;
  \item \textbf{property predicates}, which return propositions and are used in theorems.
\end{itemize}
In Lean, a \emph{predicate} is simply a definition with output type \myinline|Prop|. For example, ``this charge spectrum is complete'' or ``this spectrum allows the top Yukawa'' are predicates in exactly that sense.

Since the bounded classification problem only involves finitely many operator types, we package them into a finite formal vocabulary. At this point there are two levels of language in play: the physical operator names (\(\mu\), \(W_1\), and so on) and the Lean datatype that labels them. The role of \myinline{PotentialTerm} is simply to package those finitely many 
relevant 
operator labels into one uniform formal datatype so that the predicates below can talk about them systematically.
\begin{inputbox}{code:PotentialTerm}
\begin{codeLong}
inductive (*\physLibDocsLink[PotentialTerm]{SuperSymmetry.SU5.PotentialTerm}*)
| μ | β | Λ | W1 | W2 | W3 | W4 | K1 | K2 | topYukawa | bottomYukawa
deriving DecidableEq, Fintype
\end{codeLong}
\begin{inputboxFoot}
\footnotesize Finite formal vocabulary of operator terms used in the bounded problem.
\end{inputboxFoot}
\end{inputbox}

Once that operator vocabulary is fixed, the first basic predicate asks whether a given named term is neutral with respect to the charge data encoded by the spectrum.
\begin{inputbox}{code:AllowsTerm}
\begin{codeLong}
def (*\physLibDocsLink[AllowsTerm]{SuperSymmetry.SU5.ChargeSpectrum.AllowsTerm}*) (x : ChargeSpectrum 𝓩) (T : PotentialTerm) : Prop := 0 ∈ ofPotentialTerm x T
\end{codeLong}
\begin{inputboxFoot}
\footnotesize Neutrality predicate for a term relative to a charge spectrum.
\end{inputboxFoot}
\end{inputbox}
The top Yukawa coupling is then just one distinguished element of this finite vocabulary.

For the toy spectrum \(x_{\mathrm{toy}}\), \myinline|AllowsTerm xtoy topYukawa|  asks whether the charges of a possible top-Yukawa coupling sum to zero, so that the coupling is allowed.

The next predicate bundles the first-layer dangerous-operator exclusions into a single condition. These operators are grouped because, in the standard \(SU(5)\) interpretation, they capture the first phenomenologically dangerous channels one wants to exclude already at the charge-spectrum stage, including \(R\)-parity-violating terms and proton-decay-related couplings.
\begin{inputbox}{code:IsPhenoConstrained}
\begin{codeLong}
def (*\physLibDocsLink[IsPhenoConstrained]{SuperSymmetry.SU5.ChargeSpectrum.IsPhenoConstrained}*) (x : ChargeSpectrum 𝓩) : Prop :=
  x.AllowsTerm μ ∨ x.AllowsTerm β ∨ x.AllowsTerm Λ ∨ x.AllowsTerm W2 ∨ x.AllowsTerm W4 ∨
  x.AllowsTerm K1 ∨ x.AllowsTerm K2 ∨ x.AllowsTerm W1
\end{codeLong}
\begin{inputboxFoot}
\footnotesize Bundled first-layer dangerous-operator exclusion predicate.
\end{inputboxFoot}
\end{inputbox}
This is a good example of why it is useful to define these predicates explicitly in the API. They are not only checked in a final computation; later lemmas and reduction statements also refer to the dangerous-operator check as a single reusable condition, which one can, for example proof results about.

A third basic question is whether the charge spectrum is complete at the charge-spectrum level, meaning that both Higgs sectors and both matter sectors are actually present:
\begin{inputbox}{code:IsComplete}
\begin{codeLong}
def (*\physLibDocsLink[IsComplete]{SuperSymmetry.SU5.ChargeSpectrum.IsComplete}*) (x : ChargeSpectrum 𝓩) : Prop :=
  x.qHd.isSome ∧ x.qHu.isSome ∧ x.Q5 ≠ ∅ ∧ x.Q10 ≠ ∅
\end{codeLong}
\begin{inputboxFoot}
\footnotesize Charge-spectrum completeness predicate.
\end{inputboxFoot}
\end{inputbox}
For the toy spectrum \(x_{\mathrm{toy}}\), this predicate holds: both Higgs sectors are present and both matter sectors are nonempty. By contrast, \(y_{\mathrm{toy}}\) is not complete, because the down-type Higgs sector is absent. This is exactly the kind of simple structural statement that later becomes part of theorems about minimality and completion.

\subsection{Formalizing the bounded model space}

The bounded charge sets are not themselves a basic predicate of the API. Rather, once such bounding data are fixed, PhysLib can define the bounded class of charge spectra built from them and then state further definitions, predicates and lemmas relative to that class. Physically, this is the model space obtained once a construction tells us that only finitely many \(\overline{\mathbf 5}\)-charges and finitely many \(\mathbf{10}\)-charges are admissible.

This is what \myinline|ofFinset| does:
\begin{inputbox}{code:ofFinset}
\begin{codeLong}
def (*\physLibDocsLink[ofFinset]{SuperSymmetry.SU5.ChargeSpectrum.ofFinset}*) (S5 S10 : Finset 𝓩) : Finset (ChargeSpectrum 𝓩) :=
  let SqHd := {none} ∪ S5.map ⟨Option.some, Option.some_injective 𝓩⟩
  let SqHu := {none} ∪ S5.map ⟨Option.some, Option.some_injective 𝓩⟩
  let SQ5 := S5.powerset
  let SQ10 := S10.powerset
  (SqHd.product (SqHu.product (SQ5.product SQ10))).map toProd.symm.toEmbedding
\end{codeLong}
\begin{inputboxFoot}
\footnotesize Bounded class construction from finite allowed \(\overline{\mathbf 5}\)- and \(\mathbf{10}\)-charge menus.
\end{inputboxFoot}
\end{inputbox}
In practice, \myinline|ofFinset| is mainly used to state assumptions and definitions for the bounded problem, not as something one naively enumerates in every large case.
The definition should be read component by component.
\begin{itemize}
  \item \myinline{SqHd} is the finite set of allowed choices for the down-type Higgs sector: either it is absent (\myinline{none}) or its charge is one of the allowed \(\overline{\mathbf 5}\)-charges.
  \item \myinline{SqHu} is the analogous set of allowed choices for the up-type Higgs sector, again stored in the \(\overline{\mathbf 5}\) convention.
  \item \myinline{SQ5} is the powerset of \(S_{\bar 5}\), so it consists of all finite subsets of allowed \(\overline{\mathbf 5}\)-matter charges.
  \item \myinline{SQ10} is the powerset of \(S_{10}\), so it consists of all finite subsets of allowed \(\mathbf{10}\)-matter charges.
\end{itemize}
The Cartesian product of these four finite choice spaces therefore enumerates every charge spectrum whose Higgs and matter charges are drawn from the chosen bounded sets. In other words, \(\mathrm{ofFinset}\;S_{\bar 5}\;S_{10}\) is the exact formal analogue of the bounded class described in Section~\ref{sec:physics-to-formal}.

The toy spectrum also makes this concrete. If
\[
  S_{\bar 5}=\{-3, -2, 0\},\qquad S_{10}=\{0,-1\},
\]
then \(x_{\mathrm{toy}}\) is a member of \(\mathrm{ofFinset}\;S_{\bar 5}\;S_{10}\), since all of its Higgs and matter charges are drawn from those bounded sets. The definition \myinline|ofFinset| is therefore not an arbitrary programming helper: it is the formal object that turns the physically given charge menu into the exact bounded class over which the later theorem is stated.

Two further notions matter for the later reduction theorem.
\begin{itemize}
  \item A \myinline{Finset}, as seen above, is used for honest finite sets, where duplicates are irrelevant by design.
  \item A \myinline{Multiset} is a finite collection \emph{with} multiplicities. Multisets are useful when constructing candidate classes from several seeds or completion procedures, because the same spectrum can arise more than once before a final deduplication step.
\end{itemize}
The next two definitions implement the witness-and-completion strategy from Section~\ref{sec:physics-to-formal}. The first constructs minimal top-Yukawa seed spectra inside the bounded class; the second enlarges such seeds toward complete spectra. They are written as multiset-valued constructions because duplicates can arise before final deduplication.
\newpage
\begin{inputbox}{code:minimallyAllowsTermsOfFinset}
\begin{codeLong}
def (*\physLibDocsLink[minimallyAllowsTermsOfFinset]{SuperSymmetry.SU5.ChargeSpectrum.minimallyAllowsTermsOfFinset}*) (S5 S10 : Finset 𝓩) :
    (T : PotentialTerm) → Multiset (ChargeSpectrum 𝓩)
  | topYukawa =>
      let SqHu := S5.val
      let Q10 := toMultisetsTwo S10
      let prod := SqHu ×ˢ Q10
      let Filt := prod.filter (fun x => - x.1 + x.2.sum = 0)
      (Filt.map (fun x => ⟨none, x.1, ∅, x.2.toFinset⟩))
  -- remaining cases omitted for readability

def (*\physLibDocsLink[completionsTopYukawa]{SuperSymmetry.SU5.ChargeSpectrum.completionsTopYukawa}*) (S5 : Finset 𝓩) (x : ChargeSpectrum 𝓩) :
    Multiset (ChargeSpectrum 𝓩) :=
  (S5.val ×ˢ S5.val).map fun (qHd, q5) => ⟨qHd, x.qHu, {q5}, x.Q10⟩
\end{codeLong}
\begin{inputboxFoot}
\footnotesize Minimal top-Yukawa witnesses and their first completion step.
\end{inputboxFoot}
\end{inputbox}

The first of these definitions is a function from a potential term to the multiset of minimal witness spectra. The clause \myinline|topYukawa =>| is Lean pattern matching for the top-Yukawa case, the notation \myinline|×ˢ| denotes the relevant Cartesian product of finite collections, and the final \myinline|map| packages the resulting charge data into \myinline|ChargeSpectrum| objects. These definitions again have a direct physics reading: the first produces the minimal seed spectra that already realize the top Yukawa structure inside the bounded charge menu, while the second enlarges such a seed by adding the further Higgs and matter data needed to move toward a complete spectrum. In the last line of \myinline|completionsTopYukawa|, \myinline|x| is the seed spectrum being completed, \myinline|qHd| and \myinline|q5| are the newly chosen down-Higgs and \(\bar 5\)-matter charges, and \myinline|x.qHu| and \myinline|x.Q10| are inherited unchanged from the seed. So these are not arbitrary generators: they are formal versions of the witness-and-completion strategy introduced in Section~\ref{sec:physics-to-formal}.

\subsection{Code organization and reusability in PhysLib}

The development is meant to be read both as mathematics and as reusable code. The charge-spectrum object, its structural instances, and the first-layer predicates form a compact PhysLib API surface that later formalizations can import without re-declaring conventions. This development lives inside PhysLib not merely as a case-specific script but as part of an open-source, community-maintained library \cite{Tooby-Smith:2024vqu,PhysLibRepo,Meiburg:2025mwn}. That matters because later contributors can extend the same basic object with new predicates, helper constructions, and lemmas, rather than rebuilding the formal language from scratch.

Seen this way, the reusable layer has three parts: the core object, the predicates on that object, and the supporting lemma base that records how those predicates behave under minimality, completion, enlargement, and bounded restriction. This point matters for the next section: the main theorem is not derived from raw definitions alone, but from a stream of auxiliary lemmas that fit the vocabulary together and make later reduction arguments possible and easier.

\section{Certified Reduction: From Minimal Top-Yukawa Witnesses to Complete Charge Spectra}
\label{sec:certified-reduction}

Section~\ref{sec:formal-vocabulary} introduced the formal vocabulary. We now use it to state the main certified reduction result. Throughout this section, it is important to keep the scope precise: ``complete'' and ``viable'' refer to the \emph{charge-spectrum layer} introduced in Section~\ref{sec:physics-to-formal}. The theorem therefore classifies complete viable charge spectra in a bounded model class; it does not yet include later ingredients such as flux data, anomaly cancellation, or the absence of exotics.

By \emph{certified} we mean the following. The final candidate class is not merely produced by a search routine. Rather, membership in that class is proved inside the interactive theorem prover to be equivalent to the target physical predicate package within the bounded model class. This changes the status of the computation: the output is not just a list suggested by a script, but a formally verified classification of the complete viable charge spectra under the stated assumptions.

\subsection{Minimal top-Yukawa witnesses}

We begin with the first reduction step. A \emph{minimal top-Yukawa witness} is a charge spectrum that allows the top Yukawa term and has no proper sub-spectrum with the same property. Such witnesses isolate the smallest local charge configurations that can support the desired coupling.

In the present setting, the relevant coupling is
\[
  10\,10\,5_{H_u}.
\]
Because the implementation stores \(H_u\) in the \(\overline{\mathbf 5}\) convention, the neutrality condition takes the form
\begin{equation}
\label{eq:top-yukawa-neutrality}
  -q_{H_u} + q_{10}^{(1)} + q_{10}^{(2)} = 0.
\end{equation}

Minimality is with respect to the stored \emph{charge data}. Formally, a charge spectrum \(x\) is a minimal top-Yukawa witness if
\[
  \mathrm{AllowsTerm}\;x\;\mathrm{topYukawa}
\]
holds, but for every proper sub-spectrum \(x' \subsetneq x\), the predicate
\[
  \mathrm{AllowsTerm}\;x'\;\mathrm{topYukawa}
\]
fails. In other words, one cannot remove any Higgs or matter-charge entry from \(x\) and still retain the top Yukawa.

A small subtlety is worth stating explicitly. Since the matter sectors are stored as \myinline|Finset|s of \emph{distinct} charges, minimality is a statement about distinct charge values rather than field multiplicities. Thus a diagonal top Yukawa using the same \(\mathbf{10}\)-charge twice is represented by a witness containing that charge only once in the \(\mathbf{10}\)-sector. By contrast, an off-diagonal coupling uses two distinct \(\mathbf{10}\)-charges and therefore requires both to be present in the witness.

For example, in the toy setting of Section~\ref{sec:formal-vocabulary} with \(\mathcal{Z}=\mathbb Z\), the spectrum
\[
  x_{\mathrm{top}} :=
  \langle \mathrm{none},\ \mathrm{some}\,(-2),\ \emptyset,\ \{-1\}\rangle
\]
is a minimal top-Yukawa witness. The neutrality condition is simply
\[
  - (-2)+ (-1) + (-1) = 0,
\]
so the top Yukawa is allowed. But if one removes the Higgs-up entry or the \(\mathbf{10}\)-charge, the coupling is no longer available.

Conceptually, this is the first place where the API viewpoint pays off. Rather than beginning from all complete charge spectra, we begin from the smallest objects that already certify the local property of interest. This is encoded in \myinline|minimallyAllowsTermsOfFinset| above.

\subsection{From witnesses to complete charge spectra}

A viable complete charge spectrum must contain more than a top-Yukawa witness. It must also be complete at the charge-spectrum level and avoid the dangerous operators singled out above. The second reduction step is therefore to start from a minimal witness and study its allowed completions. This is encoded in \myinline|completionsTopYukawa| above.

If \(x\) is a minimal top-Yukawa witness and \(y\) is a viable complete charge spectrum with \(x \subseteq y\), then the formal completion routines seek an intermediate spectrum \(z\) with
\[
  x \subseteq z \subseteq y,
  \qquad \mathrm{IsComplete}\;z,
\]
constructed by adding the missing Higgs and matter sectors in a controlled way. Here ``controlled'' means that the enlargement is not arbitrary: the added sectors must still preserve the relevant viability predicates, namely absence of the selected dangerous operators and absence of their one-step Yukawa-induced regeneration.

We restate this familiar model-building picture here deliberately, because the next lemmas are best understood as its formal counterpart rather than as an unrelated proof trick.

This is exactly the same style of reasoning used in ordinary model building. One first identifies the smallest local seed carrying the desired coupling structure, and then asks how the remaining sectors can be added without spoiling the phenomenological constraints. The point of the formal proof is not to replace this physics intuition, but to make it precise enough that it becomes exhaustive.

The toy example again makes this concrete. Starting from
\[
  x_{\mathrm{top}} =
  \langle \mathrm{none},\ \mathrm{some}\,(-2),\ \emptyset,\ \{-1\}\rangle ,
\]
one possible completion step is to add a down-type Higgs of charge \(-3\) and a \(\overline{\mathbf 5}\)-matter charge \(0\), yielding our toy spectrum
\[
  x_{\mathrm{toy}} =
  \langle \mathrm{some}\,(-3),\ \mathrm{some}\,(-2),\ \{-1\},\ \{-1\}\rangle .
\]
At the charge-spectrum level this is now complete: both Higgs sectors are present and both matter sectors are nonempty. Whether such an enlargement is \emph{allowed} in the theorem depends, of course, on the viability predicates; the point of the example is only to illustrate what a completion move looks like.

This is also the sense in which the later theorem expresses \emph{closure}: once one starts from a certified seed, the permitted enlargement moves remain inside a mathematically controlled candidate class.

\subsection{Main theorem: completeness of the certified viable charge-spectrum class}

In the concrete F-theory development, the bounded input is not given directly by arbitrary finite sets \(S_{\bar 5}\) and \(S_{10}\), but by a codimension-one configuration \(I\) (following the notation used in~\cite{Krippendorf:2015kta, Lawrie:2015hia}): here this means the F-theory input record that packages the allowed \(\overline{\mathbf 5}\)- and \(\mathbf{10}\)-charge menus used to build the bounded class. We therefore write
\[
  \mathcal U(I)
  := \mathrm{ofFinset}\; I.\mathrm{allowedBarFiveCharges}\; I.\mathrm{allowedTenCharges}
\]
for the ambient bounded model class and
\[
  \mathcal V(I) := \mathrm{viableCharges}\;I
\]
for the concrete candidate class constructed in PhysLib. In what follows we will restrict to this case, however much of what we say is including in PhysLib in a reusable way for generic \(S_{\bar 5}\) and \(S_{10}\).

The central concrete theorem can then be stated as follows.

\begin{theorem}[Completeness of the certified viable charge-spectrum class]
Let \(I\) be a codimension-one configuration and let \(x \in \mathcal U(I)\). Then
\begin{equation}
\label{eq:main-completeness-v4}
  \begin{aligned}
  x \in \mathcal V(I)
  \iff\;& \mathrm{AllowsTerm}\;x\;\mathrm{topYukawa} \\
  &\wedge \neg \mathrm{IsPhenoConstrained}\;x \\
  &\wedge \neg \mathrm{YukawaGeneratesDangerousAtLevel}\;x \;1\\
  &\wedge \mathrm{IsComplete}\;x.
  \end{aligned}
\end{equation}
Equivalently, \(x \in \mathcal V(I)\) if and only if \(x\) is a viable complete charge spectrum in the bounded model class \(\mathcal U(I)\). In particular, there is no further viable complete charge spectrum in \(\mathcal U(I)\) outside the constructed class.
\end{theorem}

This is the statement that changes the interpretation of the computation. The final finite list is not merely an experimental output. It is a certified description of the entire viable class inside the chosen bounded model class.

\subsection{Proof strategy and why it matters}

The proof has four conceptual steps.
\begin{enumerate}[label=\arabic*.]
  \item Show that any spectrum allowing the top Yukawa contains a minimal top-Yukawa witness.
  \item Show that any viable complete charge spectrum containing such a witness also contains a controlled completion of that witness.
  \item Prove closure lemmas for adding allowed \(\overline{\mathbf 5}\)- and \(\mathbf{10}\)-charges while preserving the viability predicates.
  \item Conclude that every viable complete charge spectrum in the bounded model class is generated by the certified closure procedure.
\end{enumerate}

The first step isolates the local seed, the second upgrades it to a complete charge-spectrum object, and the last two show that no viable enlargement is missed. For the toy example, one should picture exactly the sequence
\[
  x_{\mathrm{top}}
  \subseteq
  x_{\mathrm{toy}}
  \subseteq
  y,
\]
where \(x_{\mathrm{top}}\) is a minimal top-Yukawa witness, \(x_{\mathrm{toy}}\) is the concrete completion introduced earlier, and \(y\) is any larger viable complete spectrum in the bounded model class. The theorem says that every such \(y\) is reached by the certified completion-and-closure mechanism.

The reason to highlight this part of the proof is that it is the bridge between physics and computation. Without it, the executable routine would just be one more search heuristic. With it, the routine becomes the implementation of a proved classification strategy. In that sense, the result resembles a generative procedure only superficially: one starts from seeds and enlarges them, but here the generation mechanism is mathematically controlled and exhaustive rather than heuristic or statistical.

A useful way to view the result is as a short \emph{lemma stream}. The early lemmas isolate minimal witnesses, the intermediate lemmas control the effect of completions, and the final closure lemmas show that the whole viable class is generated by repeated certified enlargement steps. The main theorem is the endpoint of that stream rather than a disconnected one-shot argument.

\subsection{Lean statements and proof commentary}

At the level of the formal development, the completeness result is organized in two stages. First, one proves an abstract endpoint theorem for an arbitrary multiset of charge spectra satisfying the required witness, completion, and closure hypotheses. Second, one specializes that abstract theorem to the concrete PhysLib construction \(\mathrm{viableCharges}\;I\) relevant for the \(SU(5)\) F-theory setting. This separation is conceptually useful: the first theorem captures the reusable reduction mechanism, while the second theorem packages the concrete classification used in the present case study.

The abstract endpoint theorem has the following shape:
\begin{inputbox}{code:abstractCompleteness}
\begin{codeLong}
lemma (*\physLibDocsLink[minimallyAllowsTercompleteness\_of\_isPhenoClosedQ5\_isPhenoClosedQ10]{SuperSymmetry.SU5.ChargeSpectrum.completeness_of_isPhenoClosedQ5_isPhenoClosedQ10}*)
    {S5 S10 : Finset 𝓩} {charges : Multiset (ChargeSpectrum 𝓩)}
    (charges_topYukawa : ∀ x ∈ charges, x.AllowsTerm .topYukawa)
    (charges_not_isPhenoConstrained : ∀ x ∈ charges, ¬ x.IsPhenoConstrained)
    (charges_yukawa : ∀ x ∈ charges, ¬ x.YukawaGeneratesDangerousAtLevel 1)
    (charges_complete : ∀ x ∈ charges, x.IsComplete)
    (charges_isPhenoClosedQ5 : IsPhenoClosedQ5 S5 charges)
    (charges_isPhenoClosedQ10 : IsPhenoClosedQ10 S10 charges)
    (charges_exist : ContainsPhenoCompletionsOfMinimallyAllows S5 S10 charges)
    {x : ChargeSpectrum 𝓩} (hsub : x ∈ ofFinset S5 S10) :
    x ∈ charges ↔ AllowsTerm x .topYukawa ∧
    ¬ IsPhenoConstrained x ∧ ¬ YukawaGeneratesDangerousAtLevel x 1 ∧ IsComplete x := by ...
\end{codeLong}
\begin{inputboxFoot}
\footnotesize Abstract completeness theorem for a candidate class satisfying witness, completion, and closure hypotheses.
\end{inputboxFoot}
\end{inputbox}

This statement is best read from top to bottom. The first four hypotheses say that every element of the candidate class already has the desired physical properties. The next two hypotheses express closure under admissible \(\overline{\mathbf 5}\)- and \(\mathbf{10}\)-enlargements. The final hypothesis says that the class contains the required completions of all minimal top-Yukawa witnesses. Under exactly these assumptions, membership in the candidate class is equivalent to the full target predicate inside the bounded model class.

The concrete F-theory theorem is then obtained by instantiating this abstract result with the multiset \(\mathrm{viableCharges}\;I\):
\begin{inputbox}{code:memViableChargesIff}
\begin{codeLong}
lemma (*\physLibDocsLink[mem\_viableCharges\_iff]{FTheory.SU5.ChargeSpectrum.mem_viableCharges_iff}*) {I} {x : ChargeSpectrum}
  (hsub : x ∈ ofFinset I.allowedBarFiveCharges I.allowedTenCharges) :
  x ∈ viableCharges I ↔
    AllowsTerm x topYukawa ∧
    ¬ IsPhenoConstrained x ∧
    ¬ YukawaGeneratesDangerousAtLevel x 1 ∧
    IsComplete x :=
  completeness_of_isPhenoClosedQ5_isPhenoClosedQ10
    (allowsTerm_topYukawa_of_mem_viableCharges I)
    (not_isPhenoConstrained_of_mem_viableCharges I)
    (not_yukawaGeneratesDangerousAtLevel_one_of_mem_viableCharges I)
    (isComplete_of_mem_viableCharges I)
    (isPhenoClosedQ5_viableCharges I)
    (isPhenoClosedQ10_viableCharges I)
    (containsPhenoCompletionsOfMinimallyAllows_of_subset
      (containsPhenoCompletionsOfMinimallyAllows_viableCompletions I)
      (viableCompletions_subset_viableCharges I))
    hsub
\end{codeLong}
\begin{inputboxFoot}
\footnotesize Concrete F-theory specialization identifying membership in \myinline|viableCharges I| with the target predicate package.
\end{inputboxFoot}
\end{inputbox}

This second theorem is the formal version of the completeness statement given in Section~\ref{sec:certified-reduction}. It shows that the concrete class generated in the \(SU(5)\) F-theory development is exactly the class of viable complete charge spectra in the bounded model class. In precisely that sense, \myinline|viableCharges I| should not be read as the output of a bare scan: it is accompanied by a proof that no further viable complete charge spectrum in the bounded model class has been missed.

From the physics point of view, the proof follows the same logic as the model-building strategy described above. Starting from a bounded charge spectrum satisfying the target predicates, one extracts a minimal top-Yukawa witness, upgrades it to a controlled completion, and then reconstructs the full spectrum using the closure lemmas. The Lean development therefore does not merely verify the endpoint; it mirrors the structure of the physical reasoning itself.

After the theorem statements, it is useful to see one short proof-style lemma from the same lemma stream. A basic certified fact about the concrete viable class is that every element of \myinline|viableCharges I| is complete:

\begin{inputbox}{code:isCompleteOfMemViableCharges}
\begin{codeLong}
lemma (*\physLibDocsLink[isComplete\_of\_mem\_viableCharges]{FTheory.SU5.ChargeSpectrum.isComplete_of_mem_viableCharges}*)
    (I : CodimensionOneConfig) :
    ∀ x ∈ viableCharges I, IsComplete x := by
  revert I
  decide
\end{codeLong}
\begin{inputboxFoot}
\footnotesize Short proof-style example from the certified-reduction pipeline.
\end{inputboxFoot}
\end{inputbox}

Physically, this lemma says that every spectrum in the certified viable class already contains both Higgs sectors and both matter sectors; no incomplete charge spectrum can survive into the final list. The tactic \myinline|revert I| moves the parameter \(I\) back into the goal so that the statement is presented in the right general form, and \myinline|decide| then closes the goal automatically because the completeness fact has been reduced to a decidable statement encoded in the definitions.

This completes the reduction step. Section~\ref{sec:executable-classification} turns to the executable classification built on top of this theorem and explains how the certified candidate class is turned into the final finite list of viable charge spectra.

\section{From Certified Reduction to Executable Classification}
\label{sec:executable-classification}

Section~\ref{sec:certified-reduction} established the theorem-backed reduction. The role of the present section is to explain how that proved reduction is turned into a concrete computation of the finite viable class \myinline|viableCharges I|, to separate the computed layer from the proved one, and to show how this output interfaces with a broader phenomenological workflow.

There are three levels in play throughout this discussion. First, there are the generic charge-spectrum routines in the underlying library. Second, there is their concrete specialization to a codimension-one configuration \(I\) in the \(SU(5)\) F-theory setting. Third, there are later downstream phenomenological refinements built on top of the certified charge-spectrum output. Keeping these three levels distinct helps clarify both the mathematics and the practical use of the formalization.

\subsection{From the certified reduction to a concrete closure computation}

At the generic charge-spectrum level, the computed finite class \myinline|viableCharges I| is built in three stages. First, one constructs the completed minimal witnesses: these are the smallest charge spectra that already contain the top Yukawa structure and have been completed to the charge-spectrum level while still satisfying the phenomenological exclusions. Second, one forms all admissible one-step enlargements of these spectra inside the bounded model class. Third, one iterates this enlargement step until no new viable spectra appear. Since the ambient bounded model class is finite, repeated admissible one-step enlargements must stabilize after finitely many rounds. The completeness theorem of Section~\ref{sec:certified-reduction} identifies this stabilized fixed point with the full viable class.

The first stage is implemented by the multiset
\begin{inputbox}{code:completeMinSubset}
\begin{codeLong}
def (*\physLibDocsLink[completeMinSubset]{FTheory.SU5.ChargeSpectrum.completeMinSubset}*) (S5 S10 : Finset 𝓩) : Multiset (ChargeSpectrum 𝓩) :=
  ((minimallyAllowsTermsOfFinset S5 S10 topYukawa).bind <|
      completionsTopYukawa S5).dedup.filter
    fun x => ¬ IsPhenoConstrained x ∧ ¬ YukawaGeneratesDangerousAtLevel x 1
\end{codeLong}
\begin{inputboxFoot}
\footnotesize Completed minimal witnesses used as the seed class for the executable closure routine.
\end{inputboxFoot}
\end{inputbox}
This should be read as follows. One starts from the minimal top-Yukawa witnesses inside the bounded charge menu, applies the completion routine to each of them, removes duplicates, and then filters out any spectrum that is already excluded by the phenomenological predicates. The result is therefore the finite seed class from which the later closure procedure begins.

A one-step enlargement should be understood mathematically as the passage from a given spectrum to an immediate admissible super-spectrum inside the bounded model class: one adjoins one new allowed charge datum, or missing sector datum, in such a way that the resulting spectrum still lies in the bounded model class and still satisfies the viability predicates. The helper routine \myinline|minimalSuperSet| returns exactly these admissible immediate super-spectra for a given input spectrum.

The second and third stages are implemented recursively:
\begin{inputbox}{code:viableChargesMultiset}
\begin{codeLong}
unsafe def (*\physLibDocsLink[viableChargesMultiset]{FTheory.SU5.ChargeSpectrum.viableChargesMultiset}*) (S5 S10 : Finset 𝓩) :
    Multiset (ChargeSpectrum 𝓩) := (aux (completeMinSubset S5 S10) (completeMinSubset S5 S10)).dedup
where
  /-- Auxiliary recursive function to define `viableChargesMultiset`. -/
  aux : Multiset (ChargeSpectrum 𝓩) → Multiset (ChargeSpectrum 𝓩) → Multiset (ChargeSpectrum 𝓩) :=
    fun all add =>
      /- Note that aux terminates since that every iteration the size of `all` increases,
        unless it terminates that round, but `all` is bounded in size by the number
        of allowed charges given `S5` and `S10`. -/
      if add = ∅ then all else
      let s := add.bind fun x => (minimalSuperSet S5 S10 x).val
      let s2 := s.filter fun y => y ∉ all ∧
        ¬ IsPhenoConstrained y ∧ ¬ YukawaGeneratesDangerousAtLevel y 1
      aux (all + s2) s2
\end{codeLong}
\begin{inputboxFoot}
\footnotesize Recursive executable closure routine for the certified viable class.
\end{inputboxFoot}
\end{inputbox}
Mathematically, the recursion maintains two multisets:
\begin{itemize}
  \item \texttt{all}, containing everything found so far;
  \item \texttt{add}, containing only the newly added spectra from the previous round.
\end{itemize}
From each element of \myinline|add|, the routine constructs its admissible one-step enlargements via \myinline|minimalSuperSet|. It then discards anything already seen or already ruled out by the phenomenological predicates, producing the next increment \myinline|s2|. If no new spectra appear, the recursion stops; otherwise the new spectra are added and the process repeats. For example, if \myinline|add| contains a single completed seed \(z\), then the next round examines all admissible immediate super-spectra of \(z\), discards those already seen or already excluded, and appends only the genuinely new viable spectra.

This is the executable analogue of the closure argument proved in Section~\ref{sec:certified-reduction}. The scientific point is therefore not merely that we have written a recursive routine, but that the routine is known to stabilize on a finite bounded model class and, more importantly, that its fixed point has a proved physical meaning. The keyword \myinline|unsafe| concerns the executable recursive implementation and termination packaging in Lean; it does \emph{not} weaken the mathematical certification, which comes from the separate completeness theorem.

In the concrete \(SU(5)\) F-theory development one first computes with the generic recursive routine \myinline|viableChargesMultiset S5 S10|, but the object that later appears in theorem statements is the packaged specialization \myinline|viableCharges I|. This distinction matters: the recursive multiset routine is convenient for evaluation, whereas \myinline|viableCharges I| is the theorem-facing object whose membership characterization is used in proofs.
The result \myinline|viableCharges I| explicitly contains the output of 
\myinline|viableChargesMultiset S5 S10| for the suitable \(S_5\) and \(S_{10}\).

\subsection{What is proved and what is computed}

This distinction is central to the paper and deserves explicit emphasis.
\begin{itemize}
  \item \textbf{Proved.} The semantics of the charge-spectrum object, the viability predicates, the reduction from arbitrary viable complete spectra to minimal witnesses plus controlled completions, the closure lemmas for admissible enlargements, and the completeness theorem identifying membership in \myinline|viableCharges I| with the target predicate package inside the bounded model class.
  \item \textbf{Computed.} For a chosen bounded input, evaluation produces the explicit finite multiset of viable charge spectra from which the concrete object \myinline|viableCharges I| is packaged.
\end{itemize}
The scientific claim is therefore not merely that a Lean expression evaluates to some concrete multiset \(M_I\), but that
\[
  \forall x \in \mathcal U(I),\qquad x \in M_I \leftrightarrow P_I(x),
\]
where \(P_I(x)\) denotes the conjunction of the target charge-spectrum predicates. In other words, the evaluated finite multiset has a proved semantic meaning: it is exactly the viable class at the charge-spectrum layer for the chosen bounded input.

\subsection{From charge spectra to fuller phenomenological data}

The charge-spectrum classification is only one module in a broader phenomenological pipeline. To reach fully fledged models one must typically add flux and chirality data, impose anomaly cancellation, and require an exotic-free MSSM spectrum or related low-energy conditions. In the language of the earlier F-theory analyses, this means combining the charge-spectrum object with data such as
\begin{itemize}
  \item chiralities \(M_a, M_i\),
  \item hypercharge-flux restrictions \(N_a, N_i\),
  \item anomaly cancellation constraints,
  \item and the detailed spectrum conditions defining the desired low-energy model.
\end{itemize}

The point of the present API-based organization is precisely that these later ingredients need not be mixed into the first formal object from the outset. Instead, the certified charge-spectrum classification provides an upstream reduction on which later modules can build. From the practical point of view, this means that later phenomenological layers can start from a theorem-backed reduced class rather than from the full ambient bounded model class. From the conceptual point of view, it means that one can separate the formal burden into manageable layers while preserving a clear scientific interpretation at each stage.

\subsection{Viable and non-viable regions of the bounded model class}

At the charge-spectrum layer, the elementary objects being classified are simply the bounded spectra
\[
  x \in \mathcal U(I)
\]
or, in the generic library notation, the bounded spectra in \(\mathrm{ofFinset}\;S_{\bar 5}\;S_{10}\). The completeness theorem partitions this bounded model class into two theorem-backed regions:
\begin{itemize}
  \item those spectra that lie in the certified viable class,
  \item and those spectra that do not.
\end{itemize}
It is in this precise sense that one can speak of viable and non-viable bounded charge assignments.

This matters scientifically because emptiness statements now acquire a clean interpretation. If no bounded spectrum with a given pattern of charges survives into the certified viable class, that is a mathematically justified absence result for the charge-spectrum layer. Conversely, the surviving spectra are not merely examples found by search; they are the complete bounded list of spectra satisfying the target predicate.

This is also the natural point at which numerical summaries can be reported. In a concrete bounded instance, one may report a decomposition of the ambient bounded model class into spectra that fail already at the top-Yukawa stage, spectra excluded by the phenomenological predicates, and spectra surviving to the certified viable class. Such counts are secondary to the theorem, but once the theorem is in place they become scientifically interpretable rather than merely descriptive.

\subsection{Typical workflow snippet and practical use}

This workflow snippet illustrates the practical meaning of the previous sections: the formal theorems are not merely descriptive, but directly support executable routines that return the certified viable class for a chosen bounded input. It is useful again to separate the generic library-level workflow from the concrete F-theory specialization.

At the generic charge-spectrum level one specifies bounded charge menus and then evaluates the certified viable class generated by the witness--completion--closure procedure:
\newpage
\begin{inputbox}{code:workflowGeneric}
\begin{codeLong}
-- choose bounded charge sets

def S5  : Finset Z := ...
def S10 : Finset Z := ...

-- inspect the completed minimal witnesses
#eval completeMinSubset S5 S10

-- compute the certified viable spectra in the bounded universe
#eval viableChargesMultiset S5 S10
\end{codeLong}
\begin{inputboxFoot}
\footnotesize Generic library-level workflow from bounded charge menus to the certified viable class.
\end{inputboxFoot}
\end{inputbox}

In the concrete \(SU(5)\) F-theory setting, one instead starts from a codimension-one configuration \(I\), which packages the allowed \(\overline{\mathbf 5}\)- and \(\mathbf{10}\)-charges:
\begin{inputbox}{code:workflowConfig}
\begin{codeLong}
-- choose a codimension-one configuration

def I : CodimensionOneConfig := ...

-- exactract the certified viable charge spectra for this configuration
#eval viableCharges I
\end{codeLong}
\begin{inputboxFoot}
\footnotesize Concrete configuration-level workflow for extracting \myinline|viableCharges I|.
\end{inputboxFoot}
\end{inputbox}

For a specific value of \(I\), corresponding to \myinline|.same : CodimensionOneConfig| in Lean we can give some explicit numbers. In this case 
\begin{equation*} 
S_{\bar 5}, S_{10} = \{-3, -2, -1, 0, 1, 2, 3\},
\end{equation*}
the cardinality of the ambient bounded model class \(\myinline|ofFinset _ _|\) is 1,048,576, and the cardinality of the certified viable class \(\myinline|viableCharges I|\) is 102. 

This is where the passage from theorem to practice becomes explicit. The formal definitions and theorems developed in the previous sections are not merely documentation of a search strategy; they directly support executable routines that return the finite viable class with a proved interpretation.

Once this upstream classification is in hand, a higher-level API can refine it by adding flux, chirality, or anomaly data:
\begin{inputbox}{code:workflowDownstream}
\begin{codeLong}
-- schematic downstream refinement
-- #eval viableQuanta I ...
\end{codeLong}
\begin{inputboxFoot}
\footnotesize Downstream refinement sketch built on top of the certified charge-spectrum output.
\end{inputboxFoot}
\end{inputbox}
In this way the theorem-backed charge-spectrum layer becomes a practical entry point for later phenomenological analysis rather than a disconnected formal exercise.

The overall division of labour is therefore clear. The interactive theorem prover certifies the reduction from the full bounded model class to the viable class. The executable routine enumerates that class in finite form. Later model-building layers can then use this certified output as their starting data. In this way, the formal reduction theorem becomes operational: it turns a bounded but combinatorially difficult physics question into a finite executable classification with a proved interpretation, and thereby provides certified input for later phenomenological analysis. This is the point at which the methodological difference becomes visible: comparable bounded charge classifications have previously been obtained with dedicated scan code, but here the classification is coupled to Lean proofs that certify the output under the stated assumptions.

\section{Conclusions and Outlook}

In this paper we have used a concrete \(SU(5)\) model-building problem with additional Abelian symmetries to illustrate a broader point: interactive theorem provers can be used not only to verify the endpoint of a computation, but to organize the physics problem itself into a reusable formal language. In the present case, this meant introducing a precise object language for charge spectra, formalizing the relevant phenomenological predicates, identifying minimal top-Yukawa witnesses, and proving a certified reduction from arbitrary viable complete charge spectra in a bounded model class to a finite witness--completion--closure construction. Figure~\ref{fig:workflow-certified-classification} shows a diagrammatic overview of our results.

The scientific significance of this result is not that it solves the full phenomenology problem. Rather, it shows that one can replace brute-force exploration of a combinatorially growing bounded model space by a theorem-backed classification at the charge-spectrum layer. The resulting executable routine is therefore not just a search script, but a finite computation with a proved interpretation: the output is exactly the viable complete class under the stated assumptions. In that sense, the main achievement is methodological but already scientifically meaningful. It turns a familiar model-building workflow into a form in which completeness statements, absence results, and reusable reductions can be expressed with mathematical precision.

A second lesson is that the proof strategy remains strongly guided by physics intuition. The development does not succeed by abandoning the usual model-building picture, but by formalizing it: one begins with the smallest local sectors realizing the desired coupling structure, studies their controlled completions, and proves closure under admissible enlargements. This is important conceptually. It suggests that formalization in theoretical physics need not begin from an alien language, but can often start from the same structural reasoning physicists already use when constructing and excluding models.

The case study should nevertheless be viewed as a representative proof of principle rather than a final phenomenological analysis. The completeness theorem proved in this paper concerns the charge-spectrum layer only. It does not yet fold in chirality assignments, anomaly cancellation, detailed low-energy conditions, or the full presentation of flux-related ingredients that already exist elsewhere in PhysLib but are not developed in this manuscript. The point of the present API organization is precisely that such ingredients can be attached later as further formal layers on top of a certified upstream reduction.

There are several natural next directions. One immediate expository direction is to connect the charge-spectrum classification more explicitly to the flux-related and anomaly-related components already present in PhysLib, and to invite further community contributions on top of this shared infrastructure. On the physics side, the same architecture can then be applied to broader bounded classes, including larger charge menus and more general beyond-the-Standard-Model settings where the combinatorics become even more severe. On the formal side, the present implementation already provides a reusable object language and lemma base that later formalizations can extend rather than rebuild.

A further perspective concerns AI-assisted research. None of the formal development itself depends on AI: interactive theorem proving is already a viable human workflow. The point is rather that, when AI tools are used for drafting, proposing lemmas, or exploring downstream analyses, a formal API inside an interactive theorem prover provides the stable semantics and checking layer that such tools typically lack. This suggests a division of labour in which humans or AI agents propose constructions, conjectures, reductions, or downstream analyses, while the prover enforces that the central definitions and transformations remain correct. In this way, ITP-based APIs may become a natural interface between exploratory search, executable computation, and reliable scientific reasoning.

More broadly, we hope this work helps illustrate a possible route toward scalable formal methods in theoretical physics. The value of theorem proving here is not simply that one more result has been certified. It is that a bounded but nontrivial model-building problem can be reformulated so that the combinatorics are reduced by proof, the executable search acquires a precise interpretation, and the resulting structure becomes reusable. If this strategy can be extended to richer phenomenological settings, then interactive theorem provers may become not only tools for checking arguments, but active components of how theoretical-physics workflows are designed.

\section*{Acknowledgments}

AI language tools were used for drafting and editing parts of the prose in this manuscript. The formalization, code development, and interactive-theorem-prover content were produced and checked by the authors. We thank Tyler Josephson, Andreas Schachner for discussions and the PhysLib community for discussions, infrastructure, and the reusable formal environment on which this work builds. SK has been partially supported by STFC consolidated grants ST/T000694/1 and ST/X000664/1.

\IfFileExists{utphys.bst}{\bibliographystyle{utphys}}{\bibliographystyle{unsrt}}
\bibliography{references}

\end{document}